\documentclass[ammsymb,prl,aps,twocolumn,superscriptaddress,showpacs]{revtex4}
\usepackage{amssymb}
\usepackage{graphicx}

\input{tcilatex}

\begin{document}

\title{ Dynamic phase separation in La$_{5/8-y}$Pr$_ {y}${Ca}$_{3/8}$MnO$%
_{3} $}
\author{L. Ghivelder}
\affiliation{Instituto de F\'{\i}sica, Universidade Federal do Rio de Janeiro, C.P.
68528, Rio de Janeiro, RJ 21941-972, Brazil}
\author{F. Parisi}
\affiliation{Departamento de F\'{\i}sica, Comisi\'{o}n Nacional de Energ\'{\i}a At\'{o}%
mica, Av. Gral Paz 1499 (1650) San Mart\'{\i}n, Buenos Aires, Argentina and
Escuela de Ciencia y Tecnolog\'{\i}a, UNSAM, Alem 3901, San Mart\'{\i}n,
Buenos Aires, Argentina }

\begin{abstract}
Detailed magnetization measurements in La$_{5/8-y}$Pr$_{y}${Ca}$_{3/8}$ MnO$%
_{3}$, including magnetic relaxation properties, demonstrate the dynamic
nature of the phase separated state in manganites. The difference between
the field-cooled-cooling and zero-field-cooled magnetization curves signals
the existence in the latter of blocked metastable states separated by high
energy barriers. Results of the magnetic viscosity show that the system
becomes unblocked in a certain temperature window, where large relaxation
rates are observed. We propose a simple phenomenological model in which the
system evolves through a hierarchy of energy barriers, which separates the
coexisting phases. The calculated magnetization curves using this model
reproduce all the qualitative features of the experimental data. The overall
results allowed us to construct an $H-T$ phase diagram, where frozen and
dynamic phase separation regions are clearly distinguished.
\end{abstract}

\date{\today}
\maketitle

\section{1. Introduction}

The intense investigation of rare-earth perovskite manganites, triggered by
the discovery of the well known colossal magnetoresistance (CMR) effect, has
revealed a variety of fascinating and intriguing physical properties.\cite%
{Salamon} Among these, the phenomenon known as phase separation (PS), the
coexistence at different length scales of ferromagnetic (FM) metallic and
antiferromagnetic (AFM) charge and orbital ordered insulating domains, have
recently dominated the literature on manganese oxides, and is currently
recognized as an intrinsic feature of several strongly correlated electron
systems.\cite{Dagotto} Among this class of compounds, La$_{5/8-y}$Pr$_{y}${Ca%
}$_{3/8}$MnO$_{3}$ is considered one of the prototype materials for the
study of PS. The end members of the series, La$_{5/8}$Ca$_{3/8}$MnO$_{3}$
and Pr$_{5/8}$Ca$_{3/8}$MnO$_{3}$, have a robust low temperature FM metallic
and charge ordered (CO) insulating states, respectively. The landmark paper
by Uehara and co-workers,\cite{Uehara} using magnetic, transport, and
electron microscopy techniques, showed evidence of two-phase coexistence for
intermediate Pr contents. Additional investigations including NMR,\cite%
{nmr1,nmr2} optical properties,\cite{optical} neutron scattering,\cite%
{neutron} and a variety of complementing studies,\cite{Dagotto} have
corroborated the phase separation scenario.\ Nevertheless, a clear
understanding of some basic macroscopic signatures of PS, including its
dynamic behavior, is still lacking, and the true nature of the phase
separated state is yet to be unveiled.

A relevant related issue deserving a great deal of attention nowadays is the
glassy nature of the phase separated state.\cite%
{nanoscale,Deac2001,PRLglass,PRLMathieu} The coexistence of FM and CO/AFM
phases in manganites implies the frustration of different interactions,
allowing the existence of glassy behavior. The key parameter for the
formation of the glassy state is the introduction of some kind of controlled
quenched disorder, which is able to open a window in the phase line
separating FM and CO/AF phases\cite{Aliaga,cond-mat/0302550}. Several
experimental papers have reported glassy behavior in manganites,\cite%
{Teresa,LuisY,Dho} which was attributed to cluster interaction within the
phase separated state, rather than competition between double exchange and
superexchange interactions.\cite{Deac2001,PRLglass} Glass-like dynamic
effects such as aging and rejuvenation were also found in a phase separated
manganite,\cite{PRLpablo} while a spin glass state with short range orbital
ordering but without phase separation was observed in single crystals of Eu$%
_{0.5}$Ba$_{0.5}$MnO$_{3}$.\cite{PRLMathieu} The spin dynamic of phase
separated states is commonly superposed with the growth dynamic of one phase
against the other.\cite{PRLpablo,nonvolatile} Relaxation measurements
performed on polycrystalline La$_{0.250}$Pr$_{0.375}$Ca$_{0.375}$MnO$_{3}$
revealed a high field mechanism related with the growth of the FM phase
fraction, a process that was considered as arising from a new sort of
magnetic glassiness.\cite{Deac2002} Related effects such as cooling rate
dependence on the transport and magnetic properties were also reported.\cite%
{UeharaEPL,Fischer}

In this paper we present a detailed study of the magnetic properties of
polycrystalline La$_{5/8-y}$Pr$_{y}${Ca}$_{3/8}$MnO$_{3}$ [LPCM(y)], with
emphasis on the $y$ = 0.40 sample. LPCM is one of the most studied phase
separated systems, \cite{UeharaPRL,ChemMat} and its (static) phase diagram
as a function of $y$, temperature and magnetic field was previously reported.%
\cite{Uehara,Yakubovskii} Here we focused our attention on the dynamic
properties of the phase separated state. Due to large energy barriers and
strains between the FM and CO-AFM states the system reaches low temperatures
in a highly blocked metastable state. In this context, time relaxation
measurements are important in order to reveal the equilibrium ground state.
Our results showed the existence of a temperature window where large
relaxation effects occur, and the relative fraction of the coexisting phases
rapidly changes as a function of time. We have also performed calculations
using a dynamical model, borrowed from creep theory of vortex dynamics,\cite%
{BlatterRMP} which reproduces the main features of the experimental results.
The model assumes a collective activated dynamics with diverging-like
functional form for the energy barriers. Interestingly, it predicts the
existence of multiple blocked states in the phase separated regime, arising
from the interplay between the temperature and the distance of the system to
equilibrium; the existence of these blockade states were confirmed
experimentally. Additionally, the results of magnetization as a function of
temperature ($T$) and applied field ($H$) yield an $H-T$ phase diagram of
the compound, which reveal a boundary between dynamic and frozen phase
separation effects.

\section{2. Experimental Details}

The polycrystalline samples investigated were synthesized by the liquid-mix
method starting from the metal citrates. After performing thermal treatments
at 500 $^{0}C$ for 10 hours and at 1400 $^{0}C$ for 16 hours, the obtained
powder was pressed into pellets and sintered for 2 hours at 1400 $^{0}C$.\
Scanning electron micrographs revealed a homogeneous distribution of grain
sizes, of the order of 2 $\mu $m. A single crystal with Pr content $y$ $%
\approx $ 0.375 was also investigated. Magnetization measurements were
performed with an extraction magnetometer (Quantum Design PPMS), as a
function of temperature, applied magnetic field, and elapsed time. All
temperature dependent data was measured with a cooling and warming rate of
0.8 K/min.

\section{3. Experimental Results, Discussion, and Phenomenological Model}

In order to visualize the evolution of the magnetic properties as a function
of the Pr content in the series, Fig. 1 shows the zero field-cooled
magnetization of La$_{5/8-}$$_{\mathit{y}}$Pr$_{\mathit{y}}$Ca$_{3/8}$MnO$%
_{3}$ samples, measured with $H$ = 1 T. For low Pr contents ($y$ = 0.1 and
0.2), the behavior is similar to La$_{5/8}$Ca$_{3/8}$MnO$_{3}$, with a
homogeneous FM state at low temperatures. The FM transition is shifted to
lower temperatures for increasing Pr concentrations. The $y$\ = 0.3 sample
is also a nearly homogeneous ferromagnet at low temperatures, but the
magnetization decreases through two steps when the temperature is increased.
Phase separation occurs in this system at intermediate temperatures.\cite%
{nonvolatile,MarianoPhysicaB} At the opposite end of the series, for high Pr
contents ($y$ = 0.5 and 0.625) the magnetization curves display a peak at $%
T_{CO}$ $\approx $ 230 K, interpreted as arising from the CO transition,\cite%
{PrCa95,PrCa96} and a shoulder at slightly lower temperatures, $T_{N}$ $%
\approx $ 180 K, identified through neutron data as arising from AFM order.%
\cite{neutron85} The low magnetization values indicate the existence of a
negligible amount of FM phase at low temperatures. Within the phase diagram
proposed by Dagotto and coworkers,\cite{cond-mat/0302550,Aliaga} where the
phase stability is governed by both temperature and an appropriate parameter 
\textit{g} controlling interactions, the sample with $y$\textit{\ }= 0.3 is
representative of the low \textit{g} region, with a predominantly FM
behavior at low temperatures, whereas the sample with $y$\textit{\ }= 0.5
displays features of the high \textit{g} region. It is clear from the
results plotted in Fig. 1 that the $y$ = 0.4 sample belongs to an
intermediate region, where disorder induce a \textquotedblleft
glass\textquotedblright\ state in the system. The transitions at $T_{CO}$
and $T_{N}$ are still present, but the zero field cooling magnetization show
two additional features at lower temperatures, where phase separation
phenomena are more pronounced. At very low temperatures the magnetization is
characterized by low values; an estimation based on $M$ vs. $H$ data at 2 K
yield a FM fraction of the order of 5\%. As the temperature rises this FM
fraction increases considerably at $T_{B}$ $\approx $ 23 K, a characteristic
temperature indicated in the figure, and related to the unblocking of the
low temperature frozen state,\cite{abrupt} as discussed in detail below. At $%
T_{C}$ $\approx $ 80 K this FM state becomes unstable, and the sample
changes to the antiferromagnetic state.

\begin{figure}[tbp]
\includegraphics[width=8cm,clip]{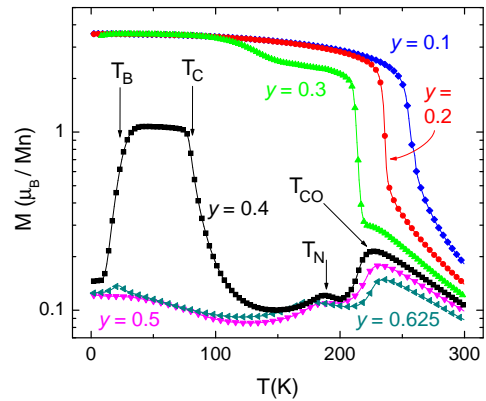}
\caption{(color online) Temperature dependence of the zero-field-cooled
magnetization of La$_{5/8-y}$Pr$_{y}$Ca$_{3/8}$MnO$_{3}$ , measured with $H$
= 1 T. For the $y$ = 0.4 sample the arrows indicate the charge-order
transition temperature ($T_{CO}$), antiferromagnetic transition ($T_{N}$),
ferromagnetic transition ($T_{C}$), and blocking temperature ($T_{B}$)}
\label{Fig1}
\end{figure}

We shall now investigate in more detail the behavior of the phase separated
state below 100 K in the $y$ = 0.4 compound. Figure 2 shows the temperature
dependence of the magnetization (a) and resistivity (b) of the LPCM(0.4)
sample, measured with $H$ = 1 T using different experimental procedures:
zero-field-cooling (ZFC), field-cooled-cooling (FCC), and field-cooled
warming (FCW). Following the FCC curve in the magnetization data, a clear FM
transition is observed at $T_{C}$ = 45 K, which is correlated with a
metal-insulator transition in the resistivity plot. Below $T_{C}$ the $M(T)$
curve changes quickly until $T$ $\sim $25K. On further cooling no changes
are observed in $M(T)$ down to the lowest temperature reached. As previously
reported,\cite{Uehara,nmr1,nmr2,optical,UeharaPRL,Podzorov} this compound
behaves as phase separated below $T_{C}$, with coexistence between the
CO-AFM and the FM phases. The magnetization value obtained at low
temperatures, $M$ $\sim $1 $\mu _{B}$/Mn, indicates that the FM fraction is
around 30\%.

\begin{figure}[tbp]
\includegraphics[width=8cm,clip]{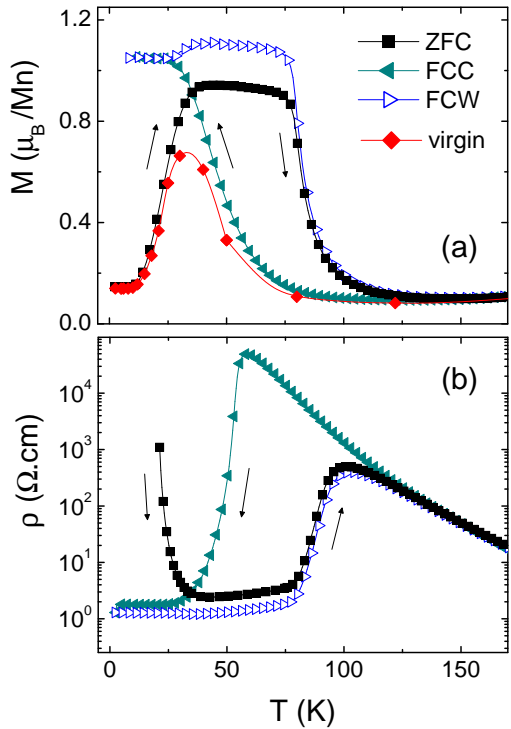}
\caption{ (color online) Magnetization (a) and resistivity (b) as a function
of temperature of La$_{0.225}$Pr$_{0.40}$Ca$_{0.375}$MnO$_{3}$. Curves
measured with zero-field-cooled, field-cooled-cooling, and
field-cooled-warming modes, with a field $H$ = 1 T. The procedure for
obtaining the virgin magnetization curve is explained in the text.}
\label{Fig2}
\end{figure}

However, the zero-field-cooled state of the sample is very different from
this picture. The results of Fig. 2 show that after zero-field cooling the
low temperature state of the system is insulating and has a very low
magnetization value, suggesting that the sample is blocked in a metastable
state with a predominance of the CO-AFM phase. Increasing the temperature in
the presence of an applied field unblocks the system, promoting a growth of
the FM phase over the AFM/CO one. The sample becomes metallic, and the
magnetization reaches and even exceeds the values obtained in the FCC
process. At higher temperatures the ZFC curve merges with the FCW one. The
FCW curve coincides with that of the FCC data until a temperature around 25
K at which an increase of the magnetization (reaching values above those of
the low temperature state) is observed; such effect is visible in several
previous investigations.\cite{UeharaPRL,Podzorov} This fact is correlated
with the decrease of the FCW resistivity curve above the reversibility
temperature.

As a way to further investigate the magnetic behavior of the system we
performed a novel experimental procedure to probe the magnetic response of
the system, which we call \textit{virgin} magnetization curve. In order to
wipe out the effect of the magnetic field on the phase separated state, the
sample is cooled without an applied field from room temperature, well within
the paramagnetic state, to a certain target temperature, then the field is
turned on to take a magnetization measurement, and subsequently the sample
is again warmed to room temperature before proceeding to the next data
point. Starting from higher temperatures, the results plotted in Fig. 2(a)
show a magnetization rise arising from the FM transition, followed by a
decrease due to the freezing of the higher temperature CO-AFM phase. The
peak in the virgin magnetization curve coincides with the temperature where
a change of behavior from metallic to insulating occurs in the FCC
resistivity.

The relaxation phenomena were investigated by measuring the magnetization as
a function of time, with $H$ = 1 T at various temperatures, after ZFC to the
desired temperature. Selected results are show in a logarithmic scale the
inset of Fig. 3. The $M(T,t)$ curves were adjusted with a logarithmic
function

\begin{figure}[tbp]
\includegraphics[width=8cm,clip]{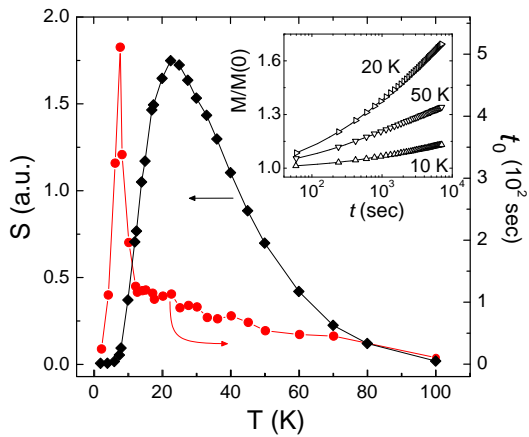}
\caption{(color online) Temperature dependence of the magnetic viscosity $S$
(left axis), and characteristic time $t_{0}$ (right axis) of La$_{0.225}$Pr$%
_{0.40}$Ca$_{0.375}$MnO$_{3}$ , obtained by fitting the time dependence of
the magnetization with Eq. (1). The inset shows the raw relaxation data,
i.e., the time evolution of the normalized magnetization, after
zero-field-cooling to selected temperatures, $T$ = 10, 20, and 50 K.}
\label{Fig3}
\end{figure}

\begin{equation}
M(T,t)=S(T)\ln (t/t_{0}+1)+M_{0}(T)  \label{1}
\end{equation}%
\newline
from which the magnetic viscosity $S$ was extracted and plotted in the main
panel of Fig. 3. At low temperatures the system is frozen in its phase
separated state, in the sense that the (low) magnetization values measured
do not evolve with time, i.e., $S(T)\approx 0$ . As the target temperature
is increased the thermal energy becomes high enough to allow to the system
to overcome the energy barriers between the coexisting phases. In this
condition the FM phase fraction shows a substantial growth as a function of
time, and the magnetic relaxation rate sharply increases. A peak in $S$ is
observed where the majority of the system becomes unblocked. At higher
temperatures the FM fraction is closer to its equilibrium value, and $S$
starts to decrease. The very large relaxation rates observed indicate that
the description of the system based on a PS state is in fact a dynamic
process, with the phase fraction of the coexisting states changing
continuously as function of time in a certain temperature window. These
relaxation measurements were also performed in a single crystal with similar
Pr content, and the same results were obtained. This confirms that the
dynamic effects observed are intrinsic to the material under investigation,
and not related to the granularity of the polycrystalline compound.

In Fig. 3 we have also plotted the temperature dependence of the macroscopic
time $t_{0}$. This parameter can be interpreted as a measurement of the time
scale at which the relaxation process occurs, taking into account that,
within an activated picture of logarithmic relaxation, the height barriers $%
U $ which can be overcome at time $t$ are $U\approx T\ln (t/t_{0}$).\cite%
{BlatterRMP} The typical values obtained for $t_{0}$, around 10$^{2}$ sec,
are orders of magnitude larger than microscopic spin flip times ($\sim $10$%
^{-12}$ sec) and even larger than relaxation times of current densities in
superconductors ($\sim $10$^{-6}$ sec) Interestingly, a sudden increase of $%
t_{0}$ is observed as $T$ is lowered, showing a cusp around 7 K. This
diverging-like behavior resembles the conventional result of the standard
theory of dynamic scaling near a phase transition,\cite{Mydosh} indicating
that a freezing process is happening.

To gain some additional insight into the low temperature behavior of the
phase separated state we developed a simple phenomenological model, which
reproduces the particular characteristics of the system. The main feature we
wish to describe is the strong blocked state that develops at low
temperatures, which is visible in both the ZFC and the FCC-FCW curves. The
model has two basic assumptions: i) the state of the system is collective;
its evolution is described as a whole in terms of a single variable, that
represents the balance between the two phases, and ii) its dynamic evolution
is hierarchical, in the sense that the most probable event happens before
the lesser probable one. Within this framework we propose a time evolution
of the system through a hierarchy of energy barriers, that applies to the
macroscopic parameters (the magnetization $M$ or, equivalently, the
ferromagnetic fraction $x$). The cooperative and hierarchical dynamics are
present in the fact that the height $U$ of the barriers are dependent of the
state of the system, i.e., $U=U(x)$. As $x$ is a macroscopic parameter, such
functional form implies that all the barriers (at the microscopic scale) of
magnitude lower than $U(x)$ were overcome before the system reaches the
state defined by the FM fraction $x$.

With these considerations, we propose a conventional activated dynamic
functional form\newline
\begin{equation}
\frac{dx}{dt}=\frac{(x_{eq}-x)}{|x_{eq}-x|}v_{0}e^{-\frac{U(x)}{T}}
\label{2}
\end{equation}%
\newline
where \textit{v}$_{0}$ represent a fixed relaxation rate, and the pre-factor
gives the sign of the time evolution, depending whether the FM fraction is
lower or higher than the equilibrium FM fraction, $x_{eq}$ ,at the given
temperature and applied field. The dependence of the energy barriers with $x$
is one of the main factors determining the dynamic characteristics of the
system. To perform the calculations we chose a diverging energy barrier
functional of the form $U(x)=U_{0}/|x_{eq}-x|$. In this way, the slow
dynamic of the system as it approaches equilibrium is guaranteed. This kind
of functional form applied for energy barriers has been extensively used to
describe vortex dynamics in high $T_{C}$ superconductors.\cite{BlatterRMP} A
simple linear form for $x_{eq}(T)$ was used, starting from $x_{eq}\ $= 0 at $%
T_{start}$ = 80 K and ending with $x_{eq}\ $=1 at $T_{end}\ $= 2.5 K. We
used $U_{0}\ $= 134.4 K, while the relaxation rate was set to \textit{v}$%
_{0}\ $= 2 sec$^{-1}$. Figure 4 shows the $x(T)$ curves obtained through the
simulated FCC, FCW and ZFC processes. A portion of the $x_{eq}(T)$ used is
also displayed in the figure, in order to proper visualize the system
behavior in respect to the equilibrium state. The FCC calculation was done
starting at temperatures above $T_{start}$ with an initial value \textit{x}
equal to zero. The temperature was lowered in steps of 0.5 K. At each
stopping temperature a \textquotedblleft measurement\textquotedblright\ was
performed, which consists in waiting a measuring time $t_{m}\ $= 60 sec
while the system is relaxing following Eq. 2. At the end of this time
period, the value $x(T)$ was obtained, which in turn was the initial $%
x(T-0.5K)$ value for the next measurement. The FCW curve was acquired in a
similar way, starting at $T_{end}$ with an initial value $x(T_{end})$ equal
to the last obtained in the FCC process. The ZFC curve was simulated
starting with a low value of $x(T_{end})$\textit{\ }\textit{= }0.02. For
completeness, we have calculated the \textit{virgin }magnetization curve,
starting at each temperature without a FM fraction, and performing the
measurement at the end of a time 5 $t_{m}$ , which is the estimated time to
measure the $M(H)$ curve until $H$ = 1 T. The resulting curve (not shown)
also reproduces very well the experimental results displayed in Fig. 2.

\begin{figure}[tbp]
\includegraphics[width=8cm,clip]{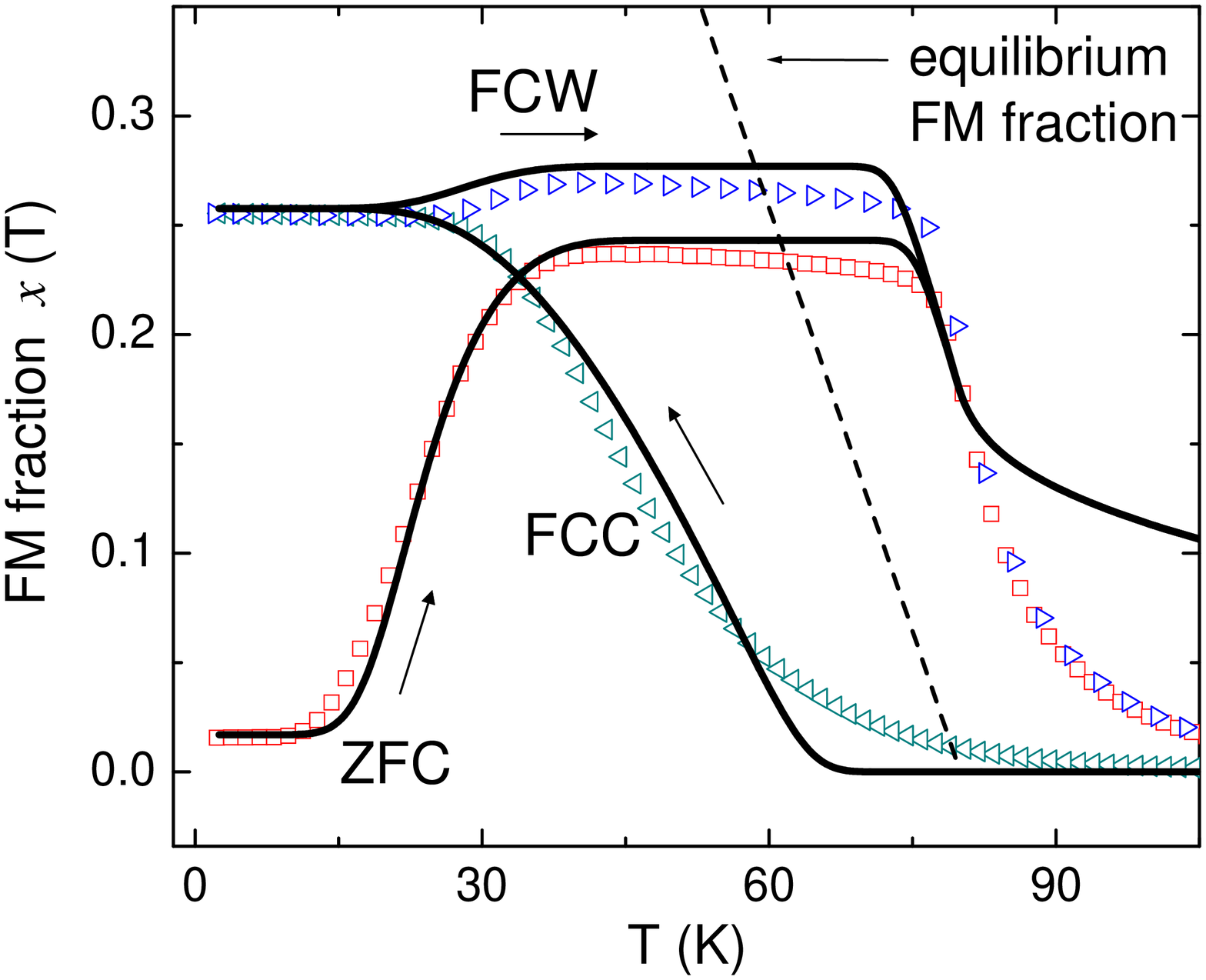}
\caption{(color online) Ferromagnetic fraction obtained from magnetization
data (symbols), compared to the calculated values (solid lines) using the
model described in the text. The dotted line is the equilibrium
ferromagnetic fraction, which reaches a value of 1.0 at low temperatures.}
\label{Fig4}
\end{figure}

The qualitatively good agreement between the observed and calculated curves
indicates that a collective mechanism is governing the dynamic evolution of
the phase separated state. This very simple model reproduces several basic
aspects of the physical response of the system, namely the existence of a
strongly blocked state at low temperatures, visible in both the ZFC and in
the FCC-FCW curves, the hysteresis observed in the whole process, the
crossing between ZFC and FCC curves, and the increase of the FCC curve above
the reversibility temperature. It is worth noting that the \textquotedblleft
reversible\textquotedblright\ behavior of the FCC-FCW curve at low
temperature is a common feature of the manganites displaying phase
separation, a singular fact not extensively discussed in the literature. In
the above described framework it is clear that the reversible behavior is
not the manifestation of an equilibrium state reached in the field cooling
process, but the collective blockade of the system. When the temperature is
high enough to unblock the system an increase of the magnetization in the
FCW curve is observed. On further heating, $x(T)$ enters in a high
temperature plateau which crosses the equilibrium curve, and gets into a
regime with an excess of FM phase. Finally, at sufficient high temperatures,
the system enters a state characterized by the quickly lost of its FM phase,
although the complete FM to non-FM phase transition on heating can not be
fully reproduced, due to the fact that the model does not allow the system
to reach the equilibrium state.

In addition, the dynamic model predicts the existence of multiple blockade
regimes. This statement lies in the fact that the effective energy barriers
distribution depends on both the temperature and the distance of the system
to the equilibrium state. A crude estimation of a blocking condition can be
made defining as blocked state that in which the FM\ fraction \textit{x }%
changes less than\textit{\ }the experimental resolution (around $10^{-5})$
within the measuring time. For the model parameters used, these
considerations lead to the condition 
\begin{equation}
T~|x_{eq}-x|~\lesssim 8K  \label{3}
\end{equation}%
for the blocking regime of the system. It is readily implied that blocked
states occur at low temperatures; within this framework the system is always
blocked below 8 K, as experimentally observed. In addition, the relation is
also satisfied in a temperature range where the FM fraction is close to its
equilibrium value. This last condition is fulfilled at temperatures close to
60 K, where $x$ reaches and eventually overcomes the equilibrium fraction
(Fig. 4). For instance, through the functional form used for $x_{eq}(T),$
equation 3 implies that with a FM fraction $x=0.25$ the system will remain
blocked in the temperature window between 45 K and 70 K. This feature is
observed as a plateau in the calculated curves of Fig. 4. In order to test
the validity of these ideas we have performed magnetization measurements
under $H=1$T on successive temperature cycles between 2 and 55 K.\ In this
way states with different values of $x$ are acceded. The obtained curve is
shown in Fig. 5, where low and high temperature blocked states can be
observed. These blocked states are characterized by the reversible behavior
of the magnetization in the upwards and downwards runs. Both blocked regimes
are separated by an unblocked region, which coincides with the temperature
range with high relaxation rates (see Fig. 3). It is readily noticed that
the temperature window in which the system is unblocked becomes narrows as
the FM fraction increase with the number of cycles, as expected from Eq. 3.

\begin{figure}[tbp]
\includegraphics[width=8cm,clip]{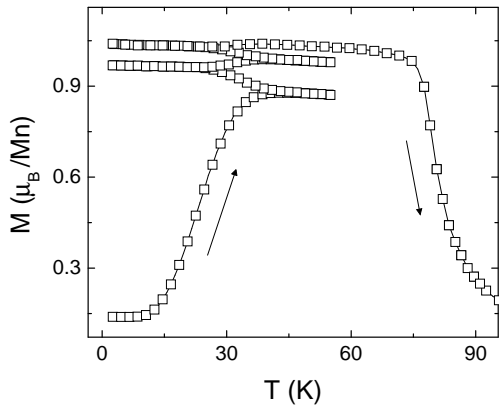}
\caption{Magnetization data, starting from zero-field-cooled data, with
several cycles in the temperature window where large relaxation effects
occur.}
\label{Fig5}
\end{figure}
\begin{figure}[tbp]
\includegraphics[width=8cm,clip]{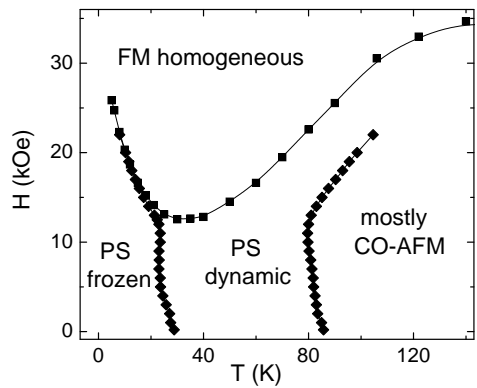}
\caption{ Zero-field-cooled $H-T\ $phase diagram of La$_{0.225}$Pr$_{0.40}$Ca%
$_{0.375}$MnO$_{3}$., The lines represent phase boundaries obtained from $M$%
\textit{\ }vs. $T$ ($\Diamond $) and $M$ vs. $H$ ($\square $) data, starting
from the ZFC state. Hysteretic effects are not displayed. The regions
depicted are homogeneous ferromagnetic at high fields, frozen phase
separation at low temperatures, dynamic phase separation at intermediate
temperatures, and mostly charge-order antiferromagnetic at higher
temperatures.}
\label{Fig6}
\end{figure}

The global results presented indicate that after zero field-cooling the
sample reaches low temperatures in a highly blocked state, with a small and
almost time independent fraction of FM phase, which can be thought as
distributed in isolated regions or clusters surrounded by a CO matrix. This
frozen-in state can be weakened, and eventually destroyed by increasing the
temperature at fixed magnetic fields, or alternatively, by increasing the
field at fixed temperatures. The latter leads to the well known metamagnetic
transition, where the entire system changes to a homogeneous FM state.
Within this context, measurements of zero filed-cooled $M$ vs. $T$ at
various fields, and $M$ vs. $H$ at various temperatures enabled us to
construct the $H-T$ phase diagram of this prototype manganite compound. The
results are shown in Fig. 6, where the different regions of the phase
diagram are depicted. It is assumed that this phase diagram refers to the
states reached after ZFC, i.e., it corresponds to the description of the
system with low initial values of the FM fraction. At high fields, above the
metamagnetic transition, the system is always in a homogeneous FM state. At
low temperatures, as already mentioned, the system is frozen in a metastable
configuration, where small FM regions are trapped in the CO-AFM background.
As the temperature is increased there is a line in the phase diagram where
the system becomes unblocked. Above this line the FM regions grow and became
the majority phase in the phase separated state. In this region the magnetic
relaxation rate is positive, and the phase separation can be viewed as
dynamic phenomena, with the relative fraction of the coexisting states
continuously changing with time. At temperatures even larger one crosses
another line related with the FM transition. In this region the FM phase is
no longer stable, and may exist solely as isolated clusters in the majority
CO-AFM matrix.

\section{4. Conclusions}

In summary, we performed an investigation of the low temperature magnetic
properties in La$_{0.225}$Pr$_{0.40}$Ca$_{0.375}$MnO$_{3}$, with emphasis on
the dynamic behavior of the phase separated state. The slow logarithmic
relaxation and the existence of field dependent blocking temperatures are
signatures that the phase separated state behaves, at least from a dynamic
point of view, as a magnetic glass. The disorder induced by chemical
substitution at the perovskite A site could be the cause of the
\textquotedblleft spread\textquotedblright\ of the free energy densities,
giving rise to a complex landscape which can be comparable to the energy
landscape in configurational space of true spin glasses. Our experimental
data shows that the dynamic of the system is better determined by the phase
competition rather than by solely magnetic interactions as in conventional
spin glasses. Such slow dynamic of the phase separated state is the main
factor determining the magnetic response of the system in different dc
experiments, and is at the basis of the well documented cooling rate
dependence of their physical response.\cite{UeharaEPL,Fischer} The fact that
the evolution of the phase separated state involves structural degrees of
freedom could be the reason for the large values of the characteristic time $%
t_{0}$ observed. The agreement between the measured magnetization curves and
the calculation performed with a model of cooperative hierarchical dynamics
with diverging barriers, and the existence of multiple blockade regimes,
gives a promising starting point to further investigate the properties of
this dynamic phase separated state.

Acknowledgements

The authors wish to thank G. Leyva for sample preparation and P. Levy for
helpful discussions. The single crystal investigated was kindly given by
S-W. Cheong. This work was partially supported by CAPES. Additional funding
came from CNPq, CONICET, and Fundaci\'{o}n Antorchas.

\end{document}